\documentstyle[preprint,prc,aps,floats,epsf]{revtex}

\preprint{IU/NTC\ \ 00--03}

\tightenlines

\begin{document}

\preprint{IU/NTC\ \ 00--13}

\title{
Effective Field Theory in\\ Nuclear Many-Body Physics}

\author{Brian D. Serot}

\address{Physics Department and Nuclear Theory Center, 
Indiana University\\
Bloomington, IN\ 47405, USA}

\author{John Dirk Walecka}

\address{Department of Physics, 
The College of William and Mary\\
Williamsburg, VA\ 23187, USA}

\maketitle

\begin{abstract}
Recent progress in Lorentz-covariant quantum field theories of the
nuclear many-body problem ({\em quantum hadrodynamics}, or QHD) is
discussed.
The importance of modern perspectives in effective field theory
and density functional theory for understanding the successes
of QHD is emphasized.

\bigskip\noindent
\ $\,${\bf To appear in:} {\it 150 Years of Quantum Many-Body Theory\/}:
A conference in honour of the 65${}^{\mathrm th}$ birthdays of
John W. Clark, Alpo J. Kallio, Manfred L. Ristig, and Sergio
Rosati.

\end{abstract}

\section{Overview}

Reference [\ref{MB7}] is a presentation entitled {\em Relativistic
Nuclear Many-Body Theory\/} given at the Seventh International Conference
on Recent Progress in Many-Body Theories held in Minneapolis, Minnesota,
in August, 1991.  This was a report on a long-term effort to understand
the nuclear many-body system in terms of relativistic quantum field
theories based on hadronic degrees of freedom \cite{Se86,IOPP}, a topic
we refer to as {\em quantum hadrodynamics}
(or QHD). An extensive, more recent review of work in this area is
contained in Ref.~[\ref{IJMPE}], and a text now exists \cite{JDW}
that provides
background material.\footnote{Extensive
references to other work in this field are contained in
Refs.~[\protect\ref{MB7}]
through
[\ref{JDW}].} 
There has been significant recent
progress in this area \cite{IJMPE,FST,EvRev,Params}, and the goal of this
contribution is to summarize briefly what has transpired since
the presentation in Ref.~[\ref{MB7}].

The only consistent framework we have for discussing the
relativistic many-body system is relativistic quantum field
theory based on a local lagrangian density.  
In any lagrangian approach,
one must first decide on the generalized coordinates, and hadronic
degrees of freedom---baryons and mesons---are the most
appropriate for ordinary nuclear systems (QHD).  
Early attempts involved
simple renormalizable models, which reproduced some basic
features of the nuclear interaction \cite{Se86}.
The advantage of such
models is that in principle, one can consistently investigate and relate
all aspects of nuclear structure to a small number of
renormalized coupling constants and masses.
The disadvantage, in addition
to the strong coupling constants that make reliable
approximation schemes difficult to come by, is that limiting the
discussion to renormalizable lagrangians is too restrictive.  Despite
these drawbacks, the simple models led to interesting insights.  In
relativistic mean-field theory (MFT), nuclear densities, the level
structure of the nuclear
shell model, and the spin dependence of nucleon--nucleus
scattering are reproduced \cite{Se86}.
The simplest model (QHD--I)
consists of baryons and isoscalar scalar and vector mesons.
A basic feature of all these models is that there are
strong scalar and vector mean fields present in the nucleus, which
cancel in the binding energy but which add to give the large spin-orbit
interaction \cite{EvRev}.

There is now overwhelming evidence that the underlying theory of the
strong interaction is {\em quantum chromodynamics} (QCD), a Yang--Mills
non-abelian gauge theory built on an internal color symmetry of a system
of quarks and gluons.  
If mass terms for the $u$ and $d$
quarks are absent in the lagrangian, QCD possesses {\em chiral symmetry}
in the nuclear domain; although spontaneously broken in manifestation, 
this symmetry should play an essential role in nuclear dynamics.  
The challenge \cite{MB7} was to understand the theoretical
basis of QHD, the successes that it had, and its limitations, in
terms of QCD.
Indeed, as we say in our summary in Ref.~[\ref{MB7}]:

\begin{quote}
\ \ \ \ \ More generally, it is probable that at low energies and large
distances, QCD can be represented by an {\em effective field theory\/}
formulated in terms of a few hadronic degrees of freedom.  All possible
couplings must be included in the low-energy effective lagrangian, which
is then to be used at tree level.  The underlying assumption of QHD is
that of a local relativistic theory formulated in terms of baryons and
the lightest mesons.  The theory is assumed to be renormalizable, and one
then attempts to extract predictions for long-range phenomena by
computing both tree-level diagrams and {\em renormalized quantum loop
corrections}.  In the end, it may turn out that this assumption is
untenable, and that the only meaningful interpretation of QHD is as an
{\em effective} theory, to be used at the tree or one-loop level.  The
limitation to renormalizable couplings may then be too restrictive.
Nevertheless, the phenomenological success of the MFT of QHD--I in the
nuclear domain implies that {\em whatever the effective field theory for
low-energy, large-distance QCD, it must be dominated by linear,
isoscalar, scalar and vector interactions}.
\end{quote}

The major progress since Ref.~[\ref{MB7}], in addition to the multitude
of applications discussed in Ref.~[\ref{IJMPE}], 
has been the following \cite{IJMPE,FST,EvRev,Params}:

\begin{itemize}

\item The understanding of QHD as a low-energy, effective lagrangian for
QCD, which can be used to improve MFT calculations systematically;

\item The understanding of the way spontaneously broken chiral symmetry
is realized in QHD;

\item The development of a consistent, controlled expansion and
approximation scheme that allows one to compute reliable results for bulk
nuclear properties;

\item The relation of relativistic MFT to density functional theory and
Kohn--Sham potentials, placing it on a sounder theoretical basis;

\item The understanding of the robustness of many of the QHD--I results.

\end{itemize}

In section 2 we discuss the relation to density functional 
theory \cite{DFT,Kohn} and Kohn--Sham potentials \cite{KS}.
\ Section 3 contains a
brief presentation of the effective lagrangian, and section 4 summarizes
some recent results.

\section{Density Functional Theory}

We begin with a discussion of nonrelativistic density functional 
theory (DFT) and generalize later to include relativity.
The basic idea behind DFT is to compute the energy $E$ of the
many-fermion system (or, at finite temperature, the grand
potential $\Omega$) as a functional of the particle density.
DFT is therefore a successor to Thomas--Fermi theory, which uses
a crude energy functional, but eliminates the need to calculate
the many-fermion wave function.

The strategy behind DFT can be seen most easily by working
in analogy to thermodynamics \cite{thermo}.
\ For a uniform system in a box of volume $V$ at temperature $T$,
one first computes the grand potential $\Omega (\mu , T, V)$, where
$\mu$ is the chemical potential.
It then follows that the number of particles $N$ is determined by
\begin{equation}
N = \langle {\hat N} \rangle = - \partial \Omega / \partial \mu 
\ .
\end{equation}
The convexity of $\Omega$ implies that $N$ is a monotonically
increasing function of $\mu$, so this relation can be inverted
for $\mu (N)$.
Finally, one makes a Legendre transformation to the Helmholtz
free energy $F(N,T,V) = \Omega (\mu (N),T,V) + \mu(N) N$ to
discuss systems with a fixed density $n = N/V$.

For a finite system, we replace the chemical potential with an
external, single-particle potential\footnote{%
In fact, one can absorb $\mu$ into the definition of $v$.  
We suppress all spin dependence at this point.}
$\sum_i v({\mathbf r}_i)$.
The grand potential is now a {\em functional}:
$\Omega ([v ({\mathbf r})], T)$, and a functional derivative
with respect to $v$ gives the particle density:\footnote{%
Higher variational derivatives yield various correlation 
functions.}
\begin{equation}
n ({\mathbf r}) = \langle {\hat n}({\mathbf r}) \rangle 
= \delta \Omega / \delta v ({\mathbf r})  \ .
\end{equation}
The convexity of $\Omega$ allows us (in principle) to invert this
relation and find $v ({\mathbf r})$ as a (complicated) functional
of $n({\mathbf r})$.
Finally, we make a functional Legendre transformation to define the
Hohenberg--Kohn free energy, which is a functional of
$n({\mathbf r})$:
\begin{equation}
F_{\mathrm {HK}} [n({\mathbf r})] = \Omega [v({\mathbf r})]
    - \int {\mathrm d}{\mathbf r} \ n({\mathbf r}) v({\mathbf r})
\ .
\end{equation}
($T$ is suppressed.)
The variational derivative of this free energy functional with
respect to $n$ now gives
\begin{equation}
\delta F_{\mathrm {HK}} / \delta n({\mathbf r})
 = - v({\mathbf r}) \ . \label{eq:varF}
\end{equation}

If we now restrict consideration to $T=0$ and $v({\mathbf r}) =0$,
then the {\em Hohenberg--Kohn theorem} follows \cite{IJMPE,Kohn}:
\ If the functional form of $F_{\mathrm {HK}}[n({\mathbf r})]$ is
known exactly, the ground-state expectation value of any
observable is a {\em unique} functional of the exact ground-state
density.
Moreover, it follows immediately from Eq.~(\ref{eq:varF}) that
the exact ground-state density can be found by minimizing the
energy functional.
Although we have assumed here that the ground state is 
non-degenerate, this assumption can be easily relaxed \cite{Kohn}.

The generalization of DFT to relativistic systems is
straightforward \cite{rel}.
The energy functional $F_{\mathrm {HK}}$ now becomes a functional
of {\em both} scalar and vector densities (or more precisely,
vector four-currents).
Extremization of the functional gives rise to variational
equations that determine the ground-state densities.

Significant progress in solving these equations was made by
Kohn and Sham \cite{KS}, who introduced a complete set of
single-particle wave functions.
In our case, these wave functions allow us to recast the variational
equations as Dirac equations for occupied orbitals.
The single-particle hamiltonian contains {\em local}, 
density-dependent, scalar and vector potentials, even when the
exact energy functional is used.
Moreover, one can introduce auxiliary (scalar and vector) fields
corresponding to the local potentials, so that the resulting
equations resemble those in a relativistic MFT 
calculation \cite{IJMPE,FST}.

The strength of the approach rests on the following 
theorem:
\begin{quote}
The exact ground-state scalar and vector densities, energy, and
chemical potential for the fully interacting many-fermion system
can be reproduced by a collection of (quasi)fermions moving in
appropriately defined, self-consistent, local, classical fields.
\end{quote}
\noindent
The proof is straightforward \cite{Kohn}.
\ Start with a collection of noninteracting fermions moving in 
an externally specified, local, one-body potential.
The exact ground state for this system is known: just calculate
the lowest-energy orbitals and fill them up.\footnote{%
For simplicity, we assume that 
the least-bound orbital is completely
filled, so the ground state is non-degenerate.}
Therefore, if one can find a suitable local, one-body potential
based on an {\em exact} energy functional, the exact
ground state of that system can be determined.
But this potential is precisely what one obtains by differentiating
the interaction parts of
$F_{\mathrm {HK}}$ with respect to $n({\mathbf r})$ \cite{Kohn}.
The resulting one-body potential will generally be density
dependent and thus must be determined self-consistently.

Several points are noteworthy.
As noted by Kohn \cite{Kohn}, the single-particle basis constructed
as described above can be considered ``density optimal'', in 
contrast to the Hartree (or Hartree--Fock) basis, which is
``total-energy optimal''.
Thus the {\em exact} scalar and vector densities are given by
sums over the squares of the Dirac wave functions, with unit
occupation probability.
Moreover, since these densities are guaranteed to make the
energy functional stationary [the external $v({\mathbf r})
= 0$], the exact ground-state energy is also obtained.
The proof that the eigenvalue of the least-bound state is
exactly the Fermi energy is given in Ref.~[\ref{eF}].
Note, however, that aside from this association, the exact
Kohn--Sham wave functions (and remaining eigenvalues) have no
known, directly observable meaning.

If one knows the exact functional form of
the energy on the density, one can describe the observables
noted in the theorem exactly (and easily) in terms of the 
Kohn--Sham basis.
Observables of this type are typically the ones calculated in
relativistic MFT.
Moreover, it has been known for many years \cite{Se86} that
the mean-field contributions dominate the single-particle
potentials at ordinary densities.
Thus, by {\em parametrizing} the energy functional in a mean-field
(or ``factorized'') form, and by fitting the parameters to
empirical bulk and single-particle nuclear data, one
should obtain an excellent approximation to the exact energy
functional in the relevant density regime.
This is the key to the success of relativistic MFT calculations,
as we will verify below, using the effective lagrangian
constructed in the next section.

\section{Effective Lagrangian}

%
\def\ua{\hbox{\b{$a$}}}
\def\uh{\hbox{\b{$h$}}}
\def\uL{\hbox{\b{$L$}}}
\def\ulambda{\hbox{\b{$\lambda$}}}
\def\uN{\!\hbox{\b{$\mkern4muN$}}}
\def\uNbar{\hbox{\b{$\mkern6mu\bar{\mkern-2muN}$}}}
\def\upi{\hbox{\b{$\mkern2mu\pi$}}}
\def\uQ{\hbox{\b{$Q$}}}
\def\uR{\hbox{\b{$R$}}}
\def\urho{\hbox{\b{$\mkern2mu\rho$}}}
\def\uU{\!\hbox{\b{$\mkern4muU$}}}
\def\uv{\hbox{\b{$v$}}}
\def\uxi{\hbox{\b{$\xi$}}}

We cannot give a detailed derivation and discussion of the effective
lagrangian of QHD in this short article, but we can illustrate the basic
principles.  To exhibit how spontaneously broken chiral symmetry is
incorporated quite generally into the hadronic theory, consider the
linear $\sigma$-model with an additional linear coupling of an isoscalar
$V^{\mu}$ to the baryon current, the so-called ``chiral $(\sigma,\omega)$
model'' \cite{JDW}.
Define right- and left-handed nucleon fields by
$\psi_{R,L} \equiv (1 \pm \gamma_5) \psi /2$ and the $SU(2)$ matrix 
$\uU \equiv \exp{( i \vec{\tau}\cdot\vec{\pi}/s_0)}$, 
where $\vec{\pi}$ is the isovector pion field. 
If $M$ is the nucleon mass, determined by the spontaneous breaking
of chiral symmetry, and $s_0= M/g_{\pi}$, then the lagrangian for the
chiral $(\sigma,\omega)$ model can be written as the $s_0 \rightarrow
\infty$ limit of the following generalized lagrangian \cite{IJMPE}
\begin{eqnarray}
         \cal{L} & = & 
  i\left[ \,
{\overline\psi}_R \gamma_{\mu} ( \partial^{\mu} 
+ i g_{\mathrm v} V^{\mu} ) \psi_R
  + {\overline\psi}_L \gamma_{\mu} ( \partial^{\mu} 
+ i g_{\mathrm v} V^{\mu} ) \psi_L
\right]                                                     \nonumber 
\\
 &&  -g_{\pi} s_0 \left(1 - \frac{\sigma}{s_0} \right) 
       \left[\,{\overline\psi}_R \uU^{\dagger} \psi_L + 
             {\overline\psi}_L \uU \psi_R \right]
+ \frac{1}{2} ( \partial_{\mu} \sigma  \partial^{\mu} \sigma )
+ \frac{1}{4} s_0^2\, {\rm tr}\, ( \partial_{\mu} \uU
\partial^{\mu} \uU^{\dagger} ) 
                                 \nonumber \\
 &&      
- {\cal V} ( \uU ,
\partial_{\mu}\uU ; \sigma)
  + \frac{1}{4}m_{\pi}^2 s_0^2 \,{\rm tr}\, (\uU +
\uU^{\dagger} - 2)
 - \frac{1}{4} F_{\mu \nu} F^{\mu \nu} 
+ \frac{1}{2} m_{\mathrm v}^2 V_{\mu} V^{\mu} \ .
                                                 \label{eq:3a}
\end{eqnarray}
For $m_{\pi}^2 = 0$, this lagrangian is evidently invariant under chiral
$SU(2)_L \times SU(2)_R$ transformations of the form
($\sigma$ and $V^{\mu}$ are unchanged)
\begin{eqnarray}
  \psi_L \rightarrow \uL \psi_L \ , \qquad
  \psi_R \rightarrow \uR \psi_R \ , \qquad
  \uU \rightarrow \uL \uU \uR^{\dagger} \ .
                                                 \label{eq:3b}
\end{eqnarray}
Here $\uL$ and $\uR$ are independent, global $SU(2)$
matrices, and the generalized potential $\cal{V}$ is chosen to be
invariant, with the limit ${\cal{V}} \rightarrow  m_{\sigma}^2
\sigma^2 /2 + O(1/s_0)$. Conventional notation is recovered with the
identification
\begin{eqnarray}
    s_0 = M / g_{\pi} \equiv f_{\pi} \ .
                                            \label{eq:3c}
\end{eqnarray}

The change of variables $\uU \equiv \uxi
\uxi, \, N_L \equiv {\uxi}^{\dagger} \psi_L, \, N_R \equiv
\uxi\psi_R$ reduces the fermion terms in the preceding lagrangian
to
\begin{eqnarray}
  \cal{L}_{\rm fermion} &=& \overline{N} \left[ i \gamma^{\mu} (
\partial_{\mu} + i \uv_{\mu} + i g_{\mathrm v} V_{\mu} ) + \gamma^{\mu}
\gamma_5 \, \ua_{\mu} - M + g_{\pi} \sigma \right]N \ ,
                                \nonumber \\
  \uv_{\mu} & \equiv & 
    - \frac{i}{2} ( \uxi^{\dagger} \partial_{\mu}
\uxi +
                    \uxi \partial_{\mu}
\uxi^{\dagger} ) \ , \qquad
     \ua_{\mu}  \equiv   
    - \frac{i}{2} ( \uxi^{\dagger} \partial_{\mu}
 \uxi -    \uxi \partial_{\mu}
\uxi^{\dagger} ) \ .
                                                     \label{eq:3d}
\end{eqnarray}
This lagrangian is invariant under the following {\em nonlinear} chiral
transformation:
\begin{equation}
      \uxi (x) \rightarrow 
      \uL \uxi (x) \uh^{\dagger}( x)
\equiv \uh (x)
   \uxi (x) \uR^{\dagger} \ , \qquad
   N(x) \rightarrow \uh (x) N(x)
            \ ,                                \label{eq:3e}
\end{equation}
where $\uh (x)$ is a local $SU(2)$ matrix. It follows
that $\uU$ still transforms globally according to 
Eq.~(\ref{eq:3b}).  
Additional mesons and interactions  can now be introduced
requiring only invariance under the {\em local} isospin 
transformations of Eq.~(\ref{eq:3e}).  
While illustrated within the framework of
a simple model, this nonlinear realization of $SU(2)_L \times
SU(2)_R$ is, in fact, quite general, and can be used as a basis for
constructing the most general QHD lagrangian \cite{IJMPE}.

The effective lagrangian, which reflects the underlying spontaneously
broken chiral symmetry of QCD, and from which the energy functional of
the previous section is obtained, is constructed from the following
series of steps \cite{IJMPE,FST}:

\begin{enumerate}

\item 
A baryon field and low-mass meson fields that concisely describe
the important interaction channels, namely, $\pi (0^-,1), \phi (0^+,0),
V_{\mu} (1^-,0)$, and $ \rho_{\mu} (1^-,1)$, are the generalized
coordinates of choice.  The pion, a Goldstone boson, is treated as in the
example above.  Higher mass meson fields are assumed to be ``integrated
out'' and their contributions contained in the effective coupling
constants.

\item 
Dimensional analysis is first used to characterize the various
terms in the effective lagrangian.  Briefly, this is done as follows. 
The initial couplings of the meson fields to the baryon fields are
linear, with a strong coupling constant $g$.  The dimensionless form of
this combination is $g \phi/M = \phi/f_{\pi}$ [see Eq.\ (\ref{eq:3c})]; 
non-Goldstone boson fields are assumed to enter in this
dimensionless form.  From the mass term of the meson fields 
$ \propto m^2 \phi^2 $, 
with $m^2 \approx M^2$, one then deduces an overall scale factor
in the lagrangian density of $f_{\pi}^2 M^2$.  From the baryon mass term
$M {\overline\psi} \psi$, one concludes that
the appropriate dimensionless form of
the baryon densities is ${\overline\psi} \psi / M f_{\pi}^2$.  
This ``naive'' dimensionless analysis 
(NDA) then implies that, after appropriate combinatorial
factors are included, the various terms in the effective lagrangian
enter with dimensionless coefficients of order unity.

\item 
The various interaction terms allowed by the $SU(2)_L \times SU(2)_R$
symmetry of QCD are then constructed using the nonlinear realization
of chiral symmetry illustrated above.  Simply writing down all possible
terms does not get one very far unless there is an {\em organizational
principle}, and the following provides the crucial insight:

\item 
Although the mean scalar and vector field energies are
{\em large} compared to the nuclear binding energy, the
dimensionless combinations 
$g_{\mathrm s}\phi_0 /M  \approx \phi_0/f_{\pi}$ and 
$g_{\mathrm v}
V_0 /M \approx V_0/f_{\pi}$ are roughly 1/3 and {\em thus provide
convenient expansion parameters}.  Furthermore, spatial variations of the
meson fields and of the baryon densities in the nucleus are observed to
occur over the scale of the nuclear surface region, and hence the
dimensionless ratio $\nabla /M$ also provides a useful expansion
quantity (as does the characterization 
of chiral symmetry violation at the
lagrangian level, $m_{\pi}/M$).

\item 
A combination of these observations allows one
to construct a hierarchy of decreasing contributions to the effective
lagrangian for the nuclear many-body system characterized by an integer
$\nu$ defined by \cite{FST}
\begin{eqnarray}
  \nu = d + \frac{n}{2} + b
 \ ,                             \label{eq:3f}
\end{eqnarray} 
where $d$ is the number of derivatives, $n$ is the number of nucleon fields,
and $b$ is the number of non-Goldstone boson fields present in the
interaction term. The effective lagrangian at various levels of $\nu$ is
given in Refs.~[\ref{IJMPE},\ref{FST}].\footnote{The extension of the
effective lagrangian to include electromagnetic interactions as an
expansion in powers of derivatives is also discussed in these references.}

\end{enumerate}

The effective lagrangian with mean meson fields 
then determines the energy
functional of the previous section, and a representation in terms of
Dirac--Hartree orbitals leads to local, nonlinear Hartree equations,
which can be solved numerically.  The extent to which nuclei exhibit this
hierarchy of interactions, and to which this effective lagrangian indeed
describes the nucleus, is discussed in the next section.

\section{Results and Summary}

\def\QED{\protect\lower0.1ex\hbox{\rule{2.2mm}{2.2mm}}}  

\begin{figure}[p]
\begin{center}
\epsfxsize=20pc 
\epsfbox{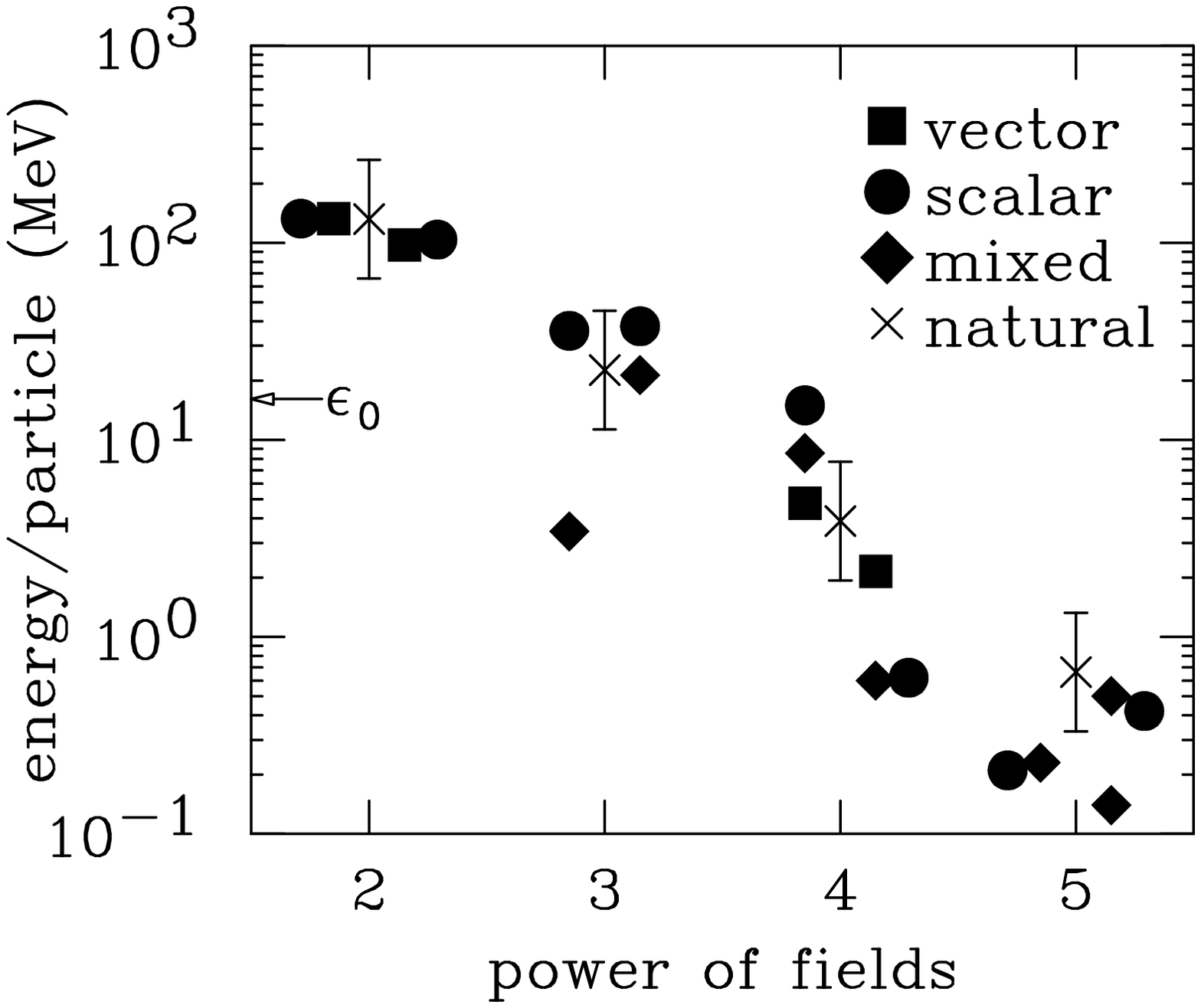} 
\vspace{.3in}
\caption{Nuclear matter energy/particle for two QHD parameter sets,
one on the left and one on the right of the error bars.
The power of fields is $b \equiv j + \ell$ for a term of the form
$(g_{\mathrm s}\phi_0)^j (g_{\mathrm v}V_0)^{\ell}$
($\ell$ is even). 
The arrow indicates the total binding energy, 
$\epsilon_{\scriptscriptstyle 0} = 16.1\,{\rm MeV}$.
Absolute values are shown.
\protect\label{fig:vmd}}
\end{center}
%
\bigskip
\bigskip
\begin{center}
\epsfxsize=20pc 
\epsfbox{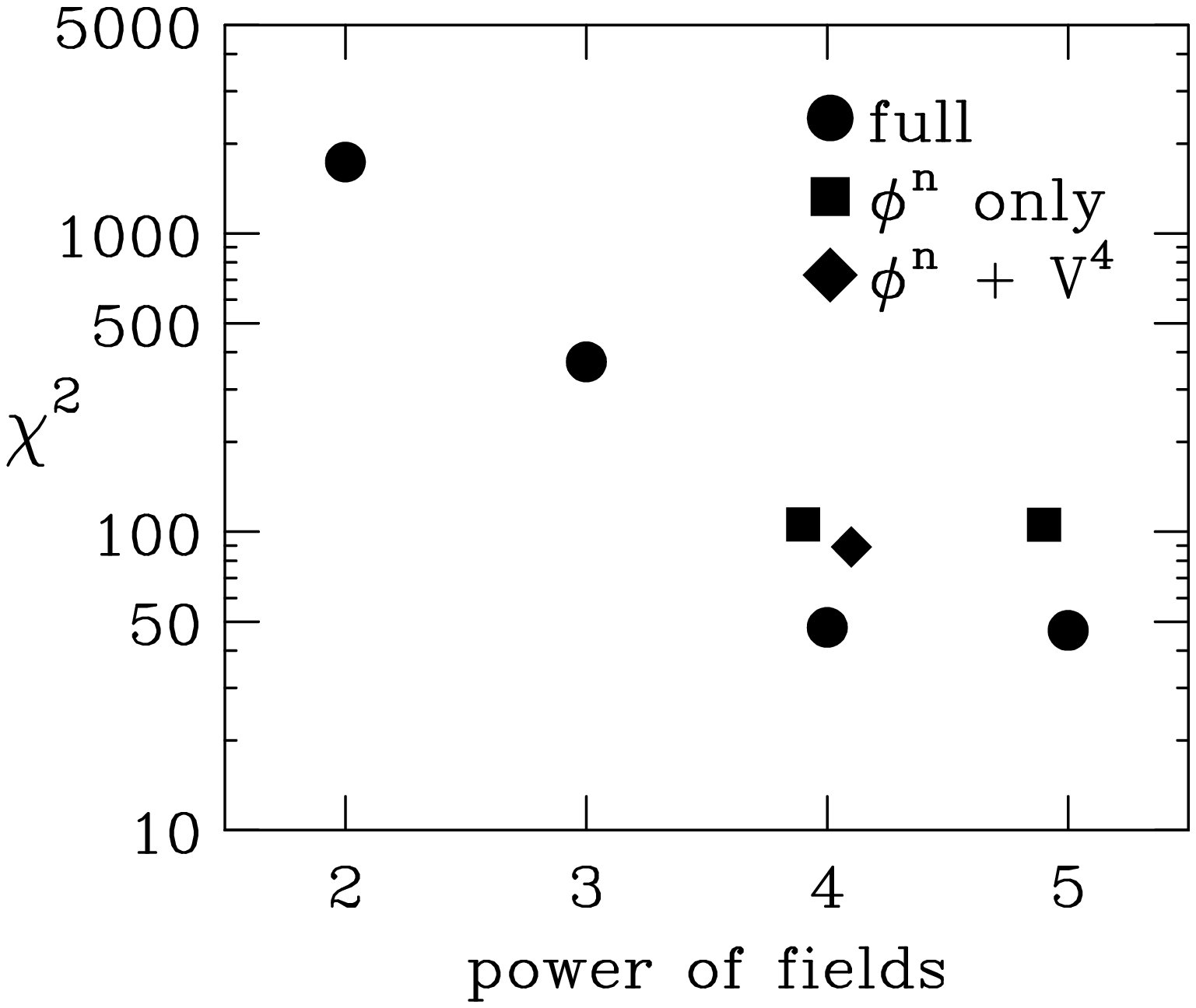} 
\vspace{.3in}
\caption{$\chi^2$ values for QHD parameter sets, as a function of the level
of truncation.  \protect\label{fig:chisq}}
\end{center}
\end{figure}

The QHD mean-field energy functional discussed above is given
in several different forms in 
Refs.~[\ref{IJMPE},\ref{FST},\ref{Params},\ref{PC}], 
as are the field equations that result from extremization.
The equations are solved self-consistently for the closed-shell
nuclei ${}^{16}$O, ${}^{40}$Ca, ${}^{48}$Ca, ${}^{88}$Sr, and
${}^{208}$Pb, and also in the nuclear matter limit.
The parameters are then best-fit to empirical
properties of the charge densities, the binding energies, and
various splittings between energy levels near the Fermi surface
using a figure of merit ($\chi^2$) defined by a weighted,
squared deviation between the 29 calculated and empirical 
values.
When working at the highest order of truncation (essentially
$\nu = 4$), the calculated results are very accurate, as we 
illustrate shortly, but they are too numerous to reproduce
here \cite{FST,PC,Params}.

The critical question is whether the hierarchal organization
of interaction terms is actually observed.
This is illustrated in Fig.~\ref{fig:vmd},
where the nuclear matter energy/particle is shown as a function
of the power of the mean fields, which is called $b$ in
Eq.~(\ref{eq:3f}).
(There are no gradient contributions in nuclear matter and
$\langle \hat{\pi} \rangle = 0$.)
The crosses and error bars are estimates based on NDA and
naturalness, that is, overall coefficients are of order unity.
It is clear that each successive term in the hierarchy is reduced
by a roughly factor of five, and thus for any reasonable desired
accuracy, the lagrangian can be truncated at a low value of $\nu$.
Derivative terms and other coupling terms are analyzed in
Ref.~[\ref{Params}], with similar conclusions.

The quality of the fits to finite nuclei
and the appropriate level of truncation
is illustrated in Fig.~\ref{fig:chisq}, where the figure of 
merit is plotted as a function of truncation order and of
various combinations of terms retained in ${\cal L}$.
The full calculations 
({\LARGE \lower0.25ex\hbox{$\bullet$}}) 
retain all allowed terms at a given
level of $\nu$, while the other two choices keep only 
the indicated subset.
There is clearly a great improvement in the fit (more than a
factor of 35)
in going from $\nu = 2$ to $\nu = 4$, but there is
no further improvement in going to $\nu = 5$.
Speaking chronologically, the $\nu = 2$ results show the level
of accuracy obtained almost 20 years ago, while the $\nu = 4$
results were obtained four years ago \cite{FST}.
\ Moreover, the $\phi^n$ only results at $\nu = 4$ 
($\,$\QED$\,$) show the
state of the situation in the late 1980s, as discussed in
Ref.~[\ref{IOPP}].
Recent work \cite{Params} shows that the full complement of
parameters at order $\nu = 4$ is underdetermined, and that only
six or seven are determined by this data set, which explains
the great success of these earlier models with a restricted
set of parameters.

In {\em summary}, 
the hadronic theory of QHD is truly a manifestation
of low-energy, strong-coupling QCD.
The modern viewpoint of QHD based on effective field theory and
density functional theory explains the accurate description of
bulk and single-particle nuclear properties.
Corrections to the mean-field parametrization of the energy
functional can be calculated systematically using the
effective hadronic lagrangian \cite{Ying}.
\ An important goal for the future is finding an efficient,
tractable, nonperturbative way to match 
the low-energy, strong-coupling, effective field theory of QHD
to the underlying QCD lagrangian.

\section*{Acknowledgements}
We thank Dick Furnstahl for useful comments.
This work was supported in part by the US Department of Energy under
contracts DE--FG02--87ER40365 and DE--FG02--97ER41023.

\end{document}